# Stickler: Defending Against Malicious CDNs in an Unmodified Browser

Amit Levy
levya@cs.stanford.edu
Stanford University
Stanford, CA

Henry Corrigan-Gibbs
henrycg@stanford.edu
Stanford University
Stanford, CA

Dan Boneh
dabo@cs.stanford.edu
Stanford University
Stanford, CA

*Abstract*—Website publishers can derive enormous performance benefits and cost savings by directing traffic to their sites through content distribution networks (CDNs). However, publishers who use CDNs today must trust their CDN not to modify the site's JavaScript, CSS, images or other media *en route* to end users. A CDN that violates this trust could inject ads into websites, downsample media to save bandwidth or, worse, inject malicious JavaScript code to steal user secrets it could not otherwise access. We present Stickler, a system for website publishers that guarantees the end-to-end authenticity of content served to end users while simultaneously allowing publishers to reap the benefits of CDNs. Crucially, Stickler achieves these guarantees *without* requiring modifications to the browser.

## I. Introduction

Transport-layer encryption (e.g., as used in HTTPS) is the industry standard for protecting the confidentiality and authenticity of data in transit on the Web. If the end user connects *directly* to a website publisher's servers then HTTPS is sufficient to authenticate the web content served over the connection. However, when there is an intermediary, such as a content distribution network (CDN), between the end user and the publisher, the HTTPS connection terminates at the CDN's servers. In this circumstance, HTTPS ensures that the *connection to the CDN* is authenticated, but it says nothing about whether the CDN is serving the publisher's intended content. Essentially, the publisher must *completely trust* the CDN to faithfully serve the site's assets to its users.

But how much should website publishers trust the CDNs that host their sites? Today, publishers have no choice but to assume that CDNs are not modifying their sites' JavaScript, images, and other assets *en route* to end users. However, this assumption is not always reasonable. For example, CDNs can inject ads in order to increase revenue, may be compelled (e.g., by powerful governments) to modify JavaScript assets to leak passwords, or might downsample image files to reduce their own bandwidth costs. Moreover, while publishers might be able to catch misbehaving CDNs by sampling CDN-served content, publishers may be unable to catch CDNs that modify content in targeted ways (e.g. only for certain IP addresses, in certain countries or after certain times).

In the last few years, a small number of CDN providers have emerged to provide content delivery services to an increasing number of websites—CloudFlare alone claims to host content for over two million websites [1]. These "consumer-grade" CDNs are much cheaper and have a much looser relationship with their customers than do traditional CDNs. Typically, website publishers sign up for an account with low-cost CDNs via a click-through web interface and often pay nothing for the service. Even though these CDNs have a tenuous business relationship with the sites they host, the sites' publishers are implicitly delegating a huge amount of trust to their CDNs: in at least one case, a no-cost CDN was able to generate CA-signed HTTPS certificates for its publishers' domains without the publishers' intervention [2]. Today, website publishers must weigh the uncertain risks of using such a service against the undeniable benefits: better availability, cheaper bandwidth, and "flash crowd" protection [3]–[8].

In this paper, we describe Stickler: a system that guarantees end-to-end integrity of website content in the face of malicious CDNs, without modifying existing browsers or CDNs. Stickler separates authenticity of the connection from authenticity of the content by signing content directly with a private key that the site owner *never* has to share with the CDN. In fact, if the site consists entirely of static content, even the publisher's *own* web servers need not hold a copy of the publisher's private signing key. End users can verify all site content independently of how they receive the pre-signed content, whether directly from the web site's servers, through a minimally trusted CDN provider, or even via a peer-to-peer CDN like CoralCDN [3].

Unlike the proposed W3C Subresource Integrity mechanism (SRI) for protecting the integrity of CDN-served content [9], Stickler does not require modifying the browser. Although browser support for integrity protection would be ideal from a performance perspective, implementing and deploying these mechanisms in commodity browsers may take years and may never reach all browser vendors and platforms. For example, Server Name Indication (SNI) was introduced in 2003 [10] and is widely used on servers to host multiple HTTPS sites on the same IP address, but did not gain widespread adoption in clients until 2011 [11] and is still not universally supported [12]. Stickler is an effective *short-term* solution for browsers that will eventually support SRI and is a useful *long-term* solution for browsers and platforms (e.g., old smartphones) that may never support SRI. We discuss SRI, the Content Security Policy mechanism [13], and other related work in more detail in Section VII.

This paper makes three main contributions:

1) **Architecture.** We introduce a CDN-compatible architecture for providing end-to-end web integrity with minimal changes to existing infrastructure.

2) **Implementation.** We present Stickler, an implementation of this architecture that works with existing servers, CDNs, and web browsers.
3) **Performance Evaluation.** We evaluate the performance of Stickler to demonstrate that it is practical and feasible to deploy on today's web.

The remainder of the paper is organized as follows: Section II presents background on CDNs and their benefits. Section III formalizes the threat model. Section IV presents an architecture that preserves end-to-end integrity of content from a malicious CDN. Section V describes our implementation of Stickler. Section VI evaluates the performance overhead of Stickler in the end user's web browser. Section VII discusses related work and Section VIII concludes.

## II. BACKGROUND

In this section, we introduce the terms used throughout this paper and explain the technical environment in which a site publisher would deploy Stickler.

### A. Participants

The four parties relevant to a Stickler deployment are: the site publisher, the content distribution network, the certificate authority, and the end user.

**Publisher.** The *site publisher* is the entity that maintains the website in question. For example, The New York Times Company is the publisher of nytimes.com. Publishers use Stickler to ensure that visitors to their site receive authentic publisher-produced content.

**Content Distribution Network (CDN).** A publisher may contract a *content distribution network* to serve the site's static web assets (images, videos, etc.) or cacheable dynamic content. Publishers typically give their CDN complete control of a specific domain on which the CDN serves the publisher's content (e.g., cdn.nytimes.com).

**Certification Authority (CA).** Publishers and CDNs can obtain signed public-key certificates from a *certification authority* to bind a domain name (e.g., nytimes.com) to a particular public key (`0xfa315...`). When a CA signs a public-key certificate, the CA is asserting that the same entity that controls the domain named in the certificate also controls the public key embedded in the certificate. Stickler only relies on HTTPS, and therefore on CAs, for reliably obtaining the publisher's public key.

**End Users.** End users access the publisher's website via their web browser, which fetches assets from both the publisher's and the CDN's servers.

### B. Benefits and Risks of CDNs

Content distribution networks (CDNs) provide web content caching as a third-party service for website publishers. CDN providers typically have edge caching servers in multiple locations around the world so that their caches are situated in geographic proximity to a site's end users. When an end user makes a web request, the CDN serves the request from its cache, otherwise it forwards the request to the publisher's server, caches the response, and returns the response to the end user.

A website publisher may choose to employ a CDN for a variety of reasons:

- CDNs can serve the bulk of a site's assets from a long-lived cache, dramatically reducing load on the publisher's servers.
- CDNs maintain edge servers around the world, so they can service cached content to clients with relatively low end-to-end latency.
- CDNs allow publishers to maintain availability in the face of a rapid spike in traffic (i.e., a "flash crowd") [14].

Assets cached by the CDN are served directly from the CDN's servers. As a result, if the end user fetches assets over HTTPS, the HTTPS connection must terminate at the CDN. The end user thus cannot distinguish between a well-behaved CDN, which caches content from the website reliably and serves it unmodified, and a malicious CDN that modifies the publisher's content.

Often, CDN-cached assets are served from a CDN-controlled domain (e.g. nytimes.procdn.biz). However, publishers may choose to have the CDN cache content on a particular subdomain (e.g., procdn.nytimes.com). Alternatively, a publisher may point the DNS records for her site to the CDN's servers to allow the CDN to serve as a front-end proxy for all requests to the site. Either way, once the website has directed the end user to retrieve content from a CDN, there is no way for the the user to verify the authenticity of that content.

## III. THREAT MODEL AND SECURITY GOALS

The goal of Stickler is to provide end-to-end integrity protection for content served to clients through a CDN. In particular, if the publisher correctly deploys Stickler, a client will only execute JavaScript code signed by the publisher and will only load publisher-signed content into the DOM, even in the presence of active attacks by the CDN. We *do not* trust the CDN for integrity. In particular, we assume that the the CDN may:

- serve stale content from the cache instead of fetching a newer version from the website.
- inject malicious client-side code into proxied responses (to sniff passwords or track clients).
- modify page content (to include advertisements).
- downsample media files (to save bandwidth)
- respond in arbitrarily malicious ways to client requests.

Providing strong integrity guarantees for CDN-hosted content is critical because integrity and confidentiality are closely linked. If the CDN can, for example, insert JavaScript code into HTML pages served to web clients, the CDN can read and exfiltrate passwords and other secret data via the client's DOM. By providing integrity guarantees, Stickler allows us to protect against confidentiality attacks that a CDN can mount by serving maliciously modified assets to the client.

We *do*, however, trust the CDN for availability: Stickler does not attempt to maintain availability when a CDN refuses to serve a publisher's content to users at all. Since CDNs have a profit motive to provide availability (though not necessarily integrity), we argue this model approximates the behavior of "consumer-grade" CDNs.

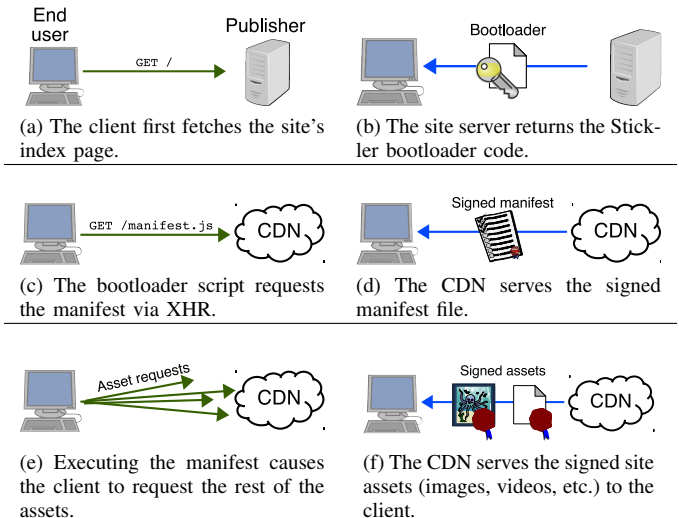

Fig. 1. The bootstrap process executed when a client visits a Stickler-protected website.

Stickler *does not* attempt to hide the web client's access patterns from the CDN. Protecting access patterns is an important, though orthogonal, problem [15]–[18]. We also *do not* attempt to protect against malicious CAs, though a publisher could deploy Stickler in parallel with other defenses to protect against CA compromise [19]–[21].

With Stickler we demonstrate that publishers can benefit from the scalability and availability benefits of a CDN without the integrity (and thus confidentiality) risks of serving assets from a third-party server.

## IV. SYSTEM ARCHITECTURE

In this section, we step through the process of loading a Stickler-protected website before examining each of the architectural components in detail.

### A. Design Overview

When a client browses to a Stickler-protected website, the DNS record for the requested domain points the client to a web server controlled by the publisher, *not* by the CDN. The client makes an HTTPS connection to this server and requests the index page for the website.

The publisher's web server returns an HTML page with the Stickler *bootloader* script embedded. This script contains (1) the publisher's public signature verification key, (2) JavaScript code to download and verify the site's assets, and (3) the location of the site's *manifest* file. Since this first request is made over an HTTPS-authenticated connection directly to the publisher's server, authenticating the connection is sufficient to authenticate the content. The publisher *never* needs to share the private-key it uses to authenticate this initial connection.

When the client executes the bootloader script, the script initiates an XHR request (i.e., AJAX request) to the CDN for the site's manifest file. Upon receiving the site manifest, the bootloader script first checks that the manifest carries a valid signature by the publisher, and then the bootloader executes the manifest file as JavaScript.

Executing the manifest file causes the client to generate a series of XHR requests to the CDN for the rest of the site's assets, each of which bears a digital signature by the publisher. When the CDN serves these assets to the client, the bootloader script verifies the publisher's signature on each asset and then processes it by invoking a function defined in the manifest. Typically, this processing just involves inserting the object into a pre-specified location in the DOM. When the user interacts with the site (e.g., by clicking a link), this could trigger more remote asset loads and signature verifications as needed to update the page content.

The publisher's server only needs to serve the initial bootloader script—the client can request all of the site's other assets directly from the CDN—and the bootloader only changes if the secret key changes, so it can be cached on the client. This means that even though requests must go directly to the publisher's server, it will only impact load times the first time a user visits the site, and should have little impact on the publisher's overall bandwidth usage. For example, in our implementation, the gzipped bootloader is 1.2 KB, meaning that a million unique visitors would require only 1.2 GB of bandwidth. At the time of writing, transferring this amount of outbound traffic from an Amazon EC2 instance costs less than $0.11 [22].

### B. Content Authentication

When the Stickler bootloader downloads an asset from the CDN, the bootloader must verify the downloaded asset (to prevent the CDN from maliciously modifying the asset). The publisher has two options for each asset. An asset can either be signed with the publisher's private key or a cryptographic hash of its contents can be embedded in the manifest. It is up to the publisher to decide which to choose.

In general, a digital signature is useful if the content is likely to change, since (1) the manifest (or whichever JavaScript file loads the asset) need not change and (2) multiple assets can exist concurrently if the CDN might serve the old (but still valid) cached assets. On the other hand, embedding the hash is suitable for content that is unlikely to ever change and where changing the asset is difficult or infeasible. For example, using a cryptographic hash to verify an asset allows it to reside on a server the publisher has no control over, such as a JavaScript or CSS library served from a third-party (e.g. Bootstrap, jQuery). Our performance evaluation (Section VI) demonstrates that there is no significant client-side performance difference between using cryptographic hashes or digital signatures for verification.

To facilitate this, Stickler requires a modification to the website authoring process. When the publisher builds their site, they must digitally sign each asset served to the client. The digital signing process can happen "on the fly" at page-load time for dynamic portions of the site's content, and can happen at "compile time" for static portions of the site's content. The requirement to sign assets is not as burdensome as it may sound: many publishers already run their HTML, JS, and CSS files through minifiers and compression tools as part of their "asset pipeline." Adding a digital signing phase to this pipeline would be relatively straightforward.

A malicious CDN might try to mount a *replay attack* by serving stale but correctly signed content to the client. For example, if a publisher discovers a security bug in one of its site's JavaScript files, the publisher would update the script file to fix the bug. Instead of serving the new script file to clients, a malicious CDN could continue to serve the validly signed but out-of-date buggy file to clients. In this way, the CDN could use a replay attack to cause a security bug in the publisher's site to persist.

To prevent replay attacks, the publisher affixes an "expiration date" to every asset before signing the asset and publishing it to the CDN. After the Stickler bootloader downloads an asset from the CDN and verifies the signature on the asset, it verifies that the asset's expiration date is in the future. By specifying short expiration dates, the publisher gains more protection from replay attacks, but must handle more traffic to their server as assets expire. This is similar to choosing a cache expiration time with the HTTP `Expires` header and, in most cases, we expect publishers to choose the same value for both.

### C. Fetching the Bootloader

As described above, the first time a client visits a Stickler-protected web site, the client connects directly to the publisher's server over HTTPS and downloads the Stickler bootloader. Since the bootloader contains the public signature verification key used to validate subsequently loaded assets, it is important that the publisher's own servers serve the bootloader to clients over an authenticated HTTPS connection.

As a result, the publisher must run a web server capable of serving the bootloader at least once to every user. This requirement is not necessarily prohibitive, though. The bootloader file is small (1.2 KB compressed) and can be cached *indefinitely*, since it only needs to change if (a) there is a bug in the bootloader or (b) the publisher wants to change their public key.

There are several ways a publisher may force the client to update a cached bootloader:

1) The publisher can serve the bootloader using the `Cache-Control: no-cache` HTTP header. Browsers still cache the content of the bootloader, but check with the publisher's server if it has changed every time the user loads the web site. In this case, the publisher's server must be able to handle a request every time each user visits the site, although unless the bootloader has changed, the response will be small (an empty 304 Not Modified response).
2) The publisher can instruct the browser (via a script in the manifest file) to clear the cached bootloader using the `window.location.reload(true)` method. This avoids making frivolous requests to the server when the bootloader hasn't changed, but won't work if the public key has changed.
3) The publisher can simply embed a new public key in the bootloader and re-sign all assets with that key. When a user first fetches an asset signed by the new key, their cached bootloader will fail to verify the asset. The bootloader may refresh itself (using `window.location.reload(true)`) to check if the publisher has embedded a new key before failing. In our prototype, we chose to implement this strategy.

### D. The Manifest File

The first file that the bootloader downloads is the site's *manifest file*, which contains a list of the site's assets. For each asset, the manifest file specifies:

- the asset's URL,
- a hash of the asset's contents (if using hash-based verification),
- an optional expiration date (to prevent the CDN from serving stale assets),
- a callback function specifying what to do with the asset once the bootloader has fetched and verified it (evaluate it as JavaScript, place its contents in the DOM, etc.), and
- a callback function specifying what to do if the asset integrity check fails.

The top-level manifest file that the bootloader fetches on site load need not specify specify every asset that the client might need. Instead, Stickler exposes a JavaScript API to the client-side application allowing it to trigger subsequent remote asset loads programmatically.

The Stickler API also allows the assets loaded in the top-level manifest file to be manifest files themselves—allowing for recursive manifest loads. This feature might be useful when different web developers are responsible for different portions of a web site—each developer could separately publish the manifest for their own portion of a site without needing to merge their changes into a top-level manifest file. Of course, to minimize the number of round-trips to the CDN servers, the application developers should minimize the depth of such a "manifest tree," since the number of round-trips increases linearly with the depth of the manifest tree.

### E. Dynamic Sites

Stickler is compatible with dynamically generated websites. There are two different methods a Stickler client-side application could use to access dynamic content.

The first method is to fetch dynamic content directly from the publisher's server over HTTPS, bypassing the CDN entirely. Since the publisher's server will be hosted on a separate domain from the CDN's servers anyways, the client can connect directly to this domain to download dynamic content. The bootloader need not verify the publisher's signature on assets fetched directly from the publisher's server. In this case, the remote peer and the content author are the same principal, so HTTPS connection-level authentication is sufficient.

The second method is to fetch dynamically generated assets via the CDN. This method might be useful if many clients will request the same dynamic asset and the CDN can cache the asset across client requests. In this case, the bootloader *does* need to verify the publisher's signature on the asset to prevent the CDN from tampering with the asset in transit. To allow for this integrity protection, the server-side application producing the dynamic asset simply must bundle the asset as a digitally signed blob that the bootloader can decode. The publisher can implement this signing-and-packaging process as part of its dynamic asset generation pipeline. The downsides of this approach are that it requires the publisher's server to digitally sign every generated object and that it requires the publisher to store its secret key online, where it may be less secure.

*F. Limitations*

Stickler has a number of important drawbacks. First, Stickler requires the publisher to sign every asset served to the client via a CDN. For some publishers, this may be relatively straightforward, but for sites with a very large number of pre-existing assets, signing every asset may be infeasible.

Second, Stickler imposes a performance penalty on the user. When visiting a Stickler-protected site, the user's browser may have to perform a large number of signature verification operations. Moreover, the user has to make an extra round-trip to the CDN to get the manifest file. Although we argue that these performance costs will not have a major impact on user experience, they may be unacceptable for some publishers. Section VI specifically evaluates the performance impact of using Stickler.

Finally, Stickler prevents (by design) a CDN from minifying script files, shrinking images, or performing other sorts of lossy compression on files without explicit publisher intervention. A publisher that heavily relies on their CDN to optimize media files and static assets values the benefit of this CDN-provided service over the risks of trusting the CDN for content integrity. However, it is important to note that Stickler does *not* interfere with normal HTTP gzip compression, since the CDN sees the content unencrypted.

## V. IMPLEMENTATION

To demonstrate that Stickler is a practical platform for building full-featured websites, we have implemented Stickler and developed a Stickler-protected demonstration website: *sticklerjs.org*. There are two components to our Stickler implementation. The first component, the *build framework*, runs on the system used to author the website. The second component, the *Stickler bootloader*, runs on the client's web browser.

**Build Framework.** The build framework runs on the development machines used to build the Stickler-protected website. The build framework takes as input (1) the publisher's secret signing key and (2) the site's static assets. The output of the build process is a set of binary blobs that the publisher serves to end users via the CDN. Each blob contains a single site asset and is either signed with the publisher's secret signing key or its hash is embedded in the site's manifest file. Our implementation allows the publisher to specify, for each asset, whether Stickler should use the signature- or hash-based technique to verify the asset's provenance. The author can additionally augment the manifest with assets from third-party sites (e.g. the jQuery CDN [23]) by specifying the URL and the expected hash of the asset in the manifest.

We implemented the build framework as a set of `node.js` [24] scripts. The framework outputs one data blob per static site asset and each blob consists of signature on the payload (the asset's binary bytes, URL path and expiration) and the payload itself. The build framework signs every blob, to allow switching between hash- and signature-based verification without rebuilding the asset blob files. Sites that already use a web framework (e.g., Django or Ruby-on-Rails) for authoring content could integrate our asset-signing process into the build pipeline used in those tools.

Stickler uses OpenSSL [25] to generate signing keys. A publisher may use any signing algorithm supported by both OpenSSL and the WebCrypto implementation. In our prototype and evaluations, we use the RSA signature algorithm with 2048-bit keys and RSA exponent `e=65537` (the default for OpenSSL). Using an RSA public key with `e=3` would decrease signature verification time, but since the cost of signature verification is small compared with the overall costs of Stickler, we chose to use OpenSSL's default RSA exponent instead. For both digital signature and hash verification we use the SHA-256 hash function.

**Stickler Bootloader.** The Stickler bootloader is the HTML document served to the web client when it makes the first request to the publisher's server. The bootloader HTML contains a copy of the publisher's public signature verification key and also contains JavaScript that loads the rest of the site. To ensure the integrity of this file, Stickler requires that this asset be served *directly from the publisher's server*, rather than the CDN. The CDN can serve all other site assets.

When executed, the bootloader's JavaScript:

1) fetches the site manifest blob file from the CDN using a cross-origin XHR request,
2) extracts the signature and payload from the blob,
3) uses the WebCrypto API [26] (or a JavaScript crypto library like SJCL [27] if WebCrypto is unsupported by the browser) to verify that the signature is valid, and
4) executes the contents of the manifest file using `eval()`.

Since the bootloader is the core of Stickler, we explain each of these steps in detail. In Step 1, the bootloader fetches the site manifest file from a URL hard-coded into the bootloader. To minimize the load on the publisher's server, the CDN hosts the manifest. Since Stickler-protected sites use a different domain for CDN-hosted content than for publisher-hosted content, this request will be a cross-origin XHR request. Although browsers prohibit such cross-origin requests by default, the website configures their assets to be served by the CDN using the CORS headers, which instruct the client's browser to allow the request [28]. Most CDNs relay any HTTP headers directly to the publisher's servers, so the publisher must simply configure *its* servers to serve cacheable assets with the appropriate CORS header.

In Step 3, the bootloader script verifies the publisher's digital signature on the payload. The script verifies the signature against a public signature verification key hard-coded into the bootloader script. If the verification check passes, the bootloader executes the content of the manifest file. If it fails, the bootloader first refreshes itself in case the publisher's verification key has changed, then shows an error to the end-user. The digital signature on the manifest file prevents a malicious CDN (or man-in-the-middle) from tricking the web client into executing arbitrary JavaScript.

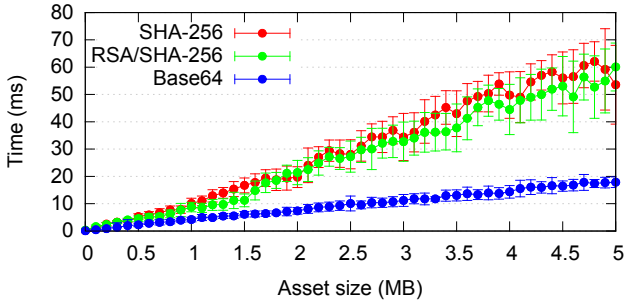

Fig. 2. The top two lines represent the time to verify an asset using digital signatures (RSA/SHA-256) and cryptographic hashes (SHA-256) in Firefox depending on the size of the asset. The difference in performance between the two methods was not statistically significant and was roughly 11 ms per 1 MB in asset size for each. The bottom line represents the additional time to encode the asset as a Base64 data-uri if it is an image or video, in order to insert it into the DOM. All experiments were run on a 3.5 Ghz quad-core Intel i7 running Firefox 35.

In Step 4, the bootloader executes the manifest file. Since the manifest file is just a JavaScript code snippet, the manifest file can itself fetch and verify site assets, which can themselves recursively fetch and verify other scripts and static resources.

Our entire bootloader is 104 lines of HTML and JavaScript code and is 1.2 KB in size when compressed. The bootloader code need not change often since the only two variables it contains are (1) the publisher's public key, and (2) the URL of the manifest file.

## VI. PERFORMANCE EVALUATION

The latency of loading an asset with Stickler is governed by three factors: network latency, cryptographic overhead (signature and hash operations), and the cost of loading the asset into the DOM. Since network latency is not particular to Stickler, our evaluation focuses on the last two factors.

Figure 2 compares the time required to perform an RSA-2048 signature verification with the time required to perform a SHA-256 hash verification. Our experiments show that for small assets, both signature and hash verification complete in under 1 ms. For larger assets, traversing the contents dominates performance and there is no measurable difference between signing and hashing. Each additional 1 MB in asset size introduces an extra 11 ms overhead for both signature and hash verification. In general, verification in either method constitutes about a 1% overhead compared to downloading that asset over a 10 Mbps network connection.

Images and other media need to further be converted to Base64 `data-uris` as an artifact of how they are embedded in the DOM. Figure 2 shows a similar experiment (using the same assets) where binary data was converted to `data-uris`. The results show that this conversion, too, is relatively small. Each additional 1 MB in asset size requires an extra 5 ms to convert it to Base64. As a concrete example, downloading a 5 MB image over a 10 Mbps connection takes approximately 4 seconds, while verifying its signature or hash and converting it to Base64 takes less than 80 ms in total.

Figure 3 compares the time it takes to render a Stickler-protected page with the time it takes to load an unprotected static HTML page as the number of assets on the page varies. In the experiment, we loaded a page with a varying number of 100 KB images (between 1 and 96) over the local network. The experiments ran on an Intel W3565 quad-core based workstation with hyperthreading running Firefox 35 and Chrome 40, and on the Google Nexus 5 mobile phone running mobile Firefox 35. As the figure demonstrates, the page render time increases roughly linearly with the number of images. Even with a relatively content-heavy page (10 MB of media assets), the page renders within 1 second on the Chrome browser and within 6 seconds on a mobile phone running Firefox. While the performance with and without Stickler on Chrome is comparable, Stickler imposes roughly a $5\times$ performance penalty when using Firefox when the number of assets on the page is large. Whether or not this overhead is acceptable will depend on the particular security requirements of the application, but we expect that for especially sensitive sites (e.g., a health data site), publishers will be willing to pay a performance cost for a security benefit.

## VII. RELATED WORK

The idea of using a small JavaScript bootloader to verify the provenance of subsequently downloaded assets has been in the Web folklore for some time and is even deployed on the Web [29], [30]. The goal of this paper is to formalize the threat model and security guarantees that such a system can provide and evaluate the performance costs of using such an architecture without modifying existing browsers.

An important piece of related work is the W3C Subresource integrity (SRI) working draft, which would augment browsers with tools to provide integrity guarantees in the face of a malicious CDN [9]. When serving a page to an SRI-enabled browser, the site author can specify the expected "fingerprint" (i.e., SHA-256 hash) of the resource to be loaded. When the browser fetches the resource, it checks the value against the expected fingerprint and only renders the object if the actual and expected fingerprints match. Browsers can enforce similar integrity checks using the "script hash" feature of Content Security Policy 2 (CSP) W3C candidate recommendation [13].

Stickler differs from these mechanisms in two important ways. First, Stickler protects content integrity *without* modifying the browser. Not needing to modify the browser makes Stickler useful as an interim solution until modern browsers support these features and as a long-term solution for browsers that may never support these features (e.g., on embedded devices). Second, Stickler allows the publisher to use *digital signatures* to verify assets, in addition to cryptographic hashes. When using digital signatures to protect content integrity, the publisher need only update the Stickler bootloader when it changes its public key. In contrast, if a publisher used only SRI to protect page integrity, the publisher would need to update the cryptographic hash in the bootloader (and potentially in many other script files as well) *every time it changed any piece of content on a page*. Use of digital signatures streamlines the page update process and moves full-site integrity protection towards practicality.

That said, it would be possible to augment the SRI and CSP recommendations to support signature-based asset integrity checks in the browser as well. One possible way to implement this feature would be to have the initial HTTPS request to the publisher's servers return an HTTP header indicating the public key that the publisher will use to sign assets served over the CDN. When the browser makes subsequent requests for site assets (images, JS files, etc.) from the CDN's servers, it would require that each response be signed with the publisher's asset signing key. The CDN could return these signatures to the end user's browser in a special HTTP response header. Although this appears to be a promising direction for content integrity protection, we leave a full security analysis of this scheme to future work.

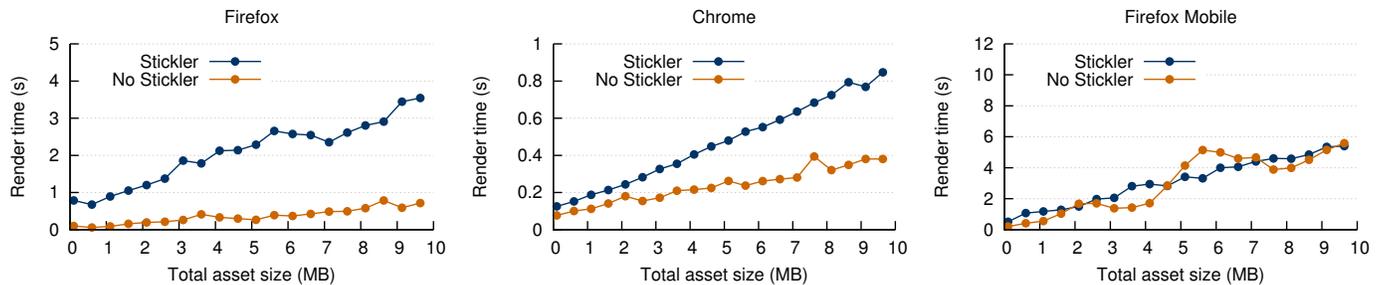

Fig. 3. Stickler imposes roughly a 5× performance overhead on Firefox, a 2× overhead on Chrome, and negligible overhead on a mobile device. In these experiments we loaded web pages into each browser while varying the number of images on the page. We measured the total time, including network latency to a server on the local host, to load the entire page. We ran the experiments on a quad-core Intel W3656 and a Google Nexus 5 mobile phone.

Liang et al. describe the weaknesses of today's "man-in-the-middle" model for using HTTPS with CDNs [31]. They propose allowing publishers to specify (in their DNS records) a set of SSL certificates that a client should accept when fetching content from the publisher's CDN. This mechanism makes it possible for the publisher to later *revoke* a CDN's certificate without having to revoke its own certificate. Unlike Stickler, the solution of Liang et al. does not protect the *integrity* of content served by a publisher's CDN.

Prior research on CDNs has focused on the design of such systems and their effectiveness at maintaining availability in the presence of "flash crowds" (traffic spikes) [3]–[8]. The Firecoral [32] system uses a browser extension to verify the integrity of content served from a peer-to-peer CDN whereas Stickler does not require modifying the browser.

Other prior work aims to protect against common browser-related security pitfalls. Bonneau's S-links make it possible to provide integrity guarantees in the presence of malicious certification authorities (CAs) [19]. Erlingsson et al. propose a system for protecting against code-injection attacks by modifying the browser to read a special server-provided security policy before processing JavaScript in the page [33]. Both of these techniques require modifying the browser and neither provides end-to-end integrity when used with a CDN.

Mylar [34] protects the confidentiality of data stored on an untrusted application server by encrypting it in the client's browser. Mylar uses a browser extension to validate the client-side code before executing it. Since we aim only to protect against a malicious CDN (*not* a malicious application server), we can avoid the use of a browser extension. Christodorescu discussed the value of using client-side cryptography to protect against server compromise [35].

## VIII. Conclusion

Stickler is a framework for protecting the integrity of a web application from a faulty or malicious CDN. Stickler uses a JavaScript-based bootloader delivered directly from the publisher's domain to verify the provenance of site assets delivered via a minimally trusted CDN. Crucially, Stickler *does not* require browser extensions or modifications and it is compatible with popular web publishing tools and techniques. Our implementation and evaluation of Stickler demonstrate its practicality and performance.

With this work, we show that website publishers can reap the manifold performance and cost benefits of using a CDN without having to put unnecessary trust in the CDN's correctness or honesty.

## Acknowledgments

We would like to thank the anonymous reviewers for their thoughtful comments. David Terei, Deian Stefan, and Keith Winstein all provided helpful suggestions on the content and presentation of the paper and, on occasion, politely listened to us rant about CDNs. This work was supported by an NDSEG fellowship, NSF, DARPA, an IARPA project provided via DoI/NBC, a grant from ONR, and by a Google faculty scholarship. Opinions, findings and conclusions or recommendations expressed in this material are those of the author(s) and do not necessarily reflect the views of DARPA or IARPA.